\documentclass[fullpage]{article}
\newcounter{mnotecount}[section]

\renewcommand{\themnotecount}{\thesection.\arabic{mnotecount}}

\newcommand{\mnote}[1]
{\protect{\stepcounter{mnotecount}}$^{\mbox{\footnotesize
$
\bullet$\themnotecount}}$ \marginpar{
\raggedright\tiny\em
$\!\!\!\!\!\!\,\bullet$\themnotecount: #1} }

\usepackage[utf8]{inputenc}
\usepackage{amsmath,amssymb}
\usepackage{amsmath}
\usepackage{bbm}
\usepackage{amsthm}
\usepackage{enumerate}
\usepackage{braket}
\usepackage{bm}
\usepackage[title]{appendix}
\usepackage{tikz}
\usepackage[none]{hyphenat}
\usepackage{geometry}
\usepackage{float}
\usepackage{fancyhdr}
\usepackage{lipsum}
\usepackage{braket}
\usepackage[labelfont=bf]{caption}
\theoremstyle{definition}

\usepackage{pifont}
\title{Positivity of Mass in Higher Dimensions}
\author{Peter Cameron\footnote{pjc96@cam.ac.uk}\;\\
Department of Applied Mathematics and Theoretical Physics\\ 
University of Cambridge\\ Wilberforce Road, Cambridge CB3 0WA, UK.}

\begin{document}
\theoremstyle{definition}
\newtheorem{definition}{Definition}[section]
\newtheorem{exmp}[definition]{Example}
\newtheorem{prop}[definition]{Proposition}
\newtheorem{lemma}[definition]{Lemma}
\newtheorem{thm}[definition]{Theorem}
\newtheorem{cor}[definition]{Corollary}
\maketitle
\begin{abstract}
\noindent 
The positive mass theorem in higher dimensions is proved using causality arguments inspired by those of Penrose, Sorkin and Woolgar \cite{PSW} in 3+1 dimensions. 
\end{abstract}
\section{Introduction}\label{Introduction}
In \cite{PSW} the authors discuss how causality arguments can be used to prove the positive mass theorem\footnote{It should be noted that this paper was never published. After discussions with people familiar with the argument and after studying it carefully myself, I believe that this was likely due to a lack of precision, rather than the methodology itself being flawed. In particular, there are issues regarding the time delay estimates in Section III. I have made every effort to be precise about the arguments used in this paper to obtain the time delay estimate Lemma \ref{lemma:timeofflight}. Specifically, I have explicitly used Minkowski retarded and advanced time co-ordinates to define a time of flight, rather than a Hamilton-Jacobi function. According to \cite{poorman} this was the source of the problems with the argument as given in \cite{PSW}. One issue which did arise and which could lead to confusion is regarding the paths of null geodesics. In \cite{PSW}, the time delay in equation (III.1.5) is calculated for a curve following the path of a Minkowski null ray and it is argued that corrections to this will be sub-leading. I have avoided making any such assumption. In fact, despite the spacetime being asymptotically flat, in positive mass Schwarzschild in 3+1 dimensions, null geodesics arbitrarily far from the mass source eventually diverge infinitely far from Minkowski geodesics. This is described in more detail in \cite{Penrose}. In fact, Lemma \ref{lemma:compactification} proves that this does not happen for a certain class of spacetimes in higher dimensions. This is then used in Lemma \ref{lemma:timeofflight} to derive a time delay estimate in a more precise way.}. Their arguments have the advantage of relying specifically on properties of spacetime which one would expect to be characteristic of positive mass, in particular the focusing and retarding of null geodesics which pass near the source (as well as the corresponding relative time advancement of geodesics passing far from the source). Conversely, null geodesics which pass far from a negative mass source will be delayed relative to those which pass nearby. This is used to argue that a certain class of negative mass spacetimes must contain a null line as defined by Galloway \cite{Galloway}:
\begin{definition}\label{defn:nullline}
A null line is an inextendible null geodesic which is achronal. In particular, null lines cannot contain conjugate points. 
\end{definition}

On the other hand, under certain conditions (see for example Theorem \ref{thm:Borde} in Section \ref{A Focusing Theorem in Higher Dimensions}) it is possible to prove that such a null line cannot exist. This allows us to prove a positive mass theorem among spacetimes satisfying these conditions (Corollary \ref{cor}). If the Einstein equations are assumed, conditions on the Riemann tensor can equivalently be stated as conditions on the stress-energy tensor and hence can be thought of as requirements on the matter content of spacetime. 

The argument given in \cite{PSW} concerns only $3+1$ dimensional spacetimes and relies on the fact that null geodesics become infinitely delayed, relative to those nearer to a negative mass source, as we let their distance of closest approach tend to infinity (see \cite{Penrose} for further discussion). As discussed in \cite{PCMD} for the specific example of the Schwarzschild metric, this effect becomes vanishingly small in higher dimensions (i.e. in $D+1$ dimensions with $D\geq4$), owing to the faster decay at infinity of the leading order non-Minkowski terms in the metric. Indeed, \cite{PSW} contains a brief discussion of this point and notes that it prevents the argument given from immediately generalising to higher dimensions.

However, in \cite{poorman2} the result is generalised to include higher dimensions (albeit with a more restrictive class of metrics than those considered in \cite{PSW}). The assumptions made in this paper are weaker than those of \cite{poorman2} - namely we will require the metric to be \textit{uniformly Schwarzschildean} (Definition \ref{defn:asymptoticallyschwarzschildean}) as in \cite{poorman} rather than the more restrictive assumption that it is \textit{strongly uniformly Schwarzschildean} (\cite{poorman2} Section 1). We will also drop the assumption in \cite[Theorem 1.1]{poorman2} that the spacetime is weakly asymptotically regular\footnote{A spacetime is weakly asymptotically regular if every null line starting in the domain of outer communications either crosses an event horizon or reaches arbitrarily large values of $r$ in the asymptotically flat regions.}). On the other hand, our proof will apply only to higher dimensions. We will also be unable to obtain a rigidity result in the $m=0$ case. We hope that this method of proof will be enlightening as it is closer in spirit to the methods employed in \cite{PSW}, namely the construction of a null line as a \textit{fastest causal curve} (Section \ref{Constructing a Fastest Causal Curve}) between two generators of past and future null infinity.

The theorem we will prove is the following (see later sections for definitions of various terms):
\begin{thm}\label{thm:nullline}
Let $(M,g)$ be a uniformly Schwarzschildean spacetime in $D+1$ dimensions ($D\geq4$) with ADM mass $m_{ADM}<0$ and suppose $\mathcal{D}\cup\mathcal{I}$ is a globally hyperbolic subset of $\tilde{M}$. Then $(M,g)$ contains a null line.
\end{thm}

For $3+1$ dimensional spacetimes, the argument in \cite{PSW}, which we now outline, proceeds by constructing a \textit{fastest causal curve} between a generator, $\Lambda^-$, of past null infinity (denoted $\mathcal{I}^-$) and a generator, $\Lambda^+$, of future null infinity (denoted $\mathcal{I}^+$). A causal (i.e. nowhere spacelike) curve is said to be \textit{faster} than some other if it departs $\mathcal{I}^-$ no earlier and arrives at $\mathcal{I}^+$ no later. A fastest causal curve (if it exists) is defined to be such that no other causal curve is faster. 

In order to specify the generator $\Lambda^+$, we specify some radial, outgoing, future pointing, Minkowski-null direction\footnote{We follow Penrose's ``abstract index notation'' where Latin indices label the rank of a tensor, while Greek indices denote the components of a tensor in some co-ordinate basis.} $k^a$. We then define $\Lambda^+$ to be the intersection of $\mathcal{I}^+$ with the union of all null geodesics whose future pointing tangent vector asymptotes towards this direction at future null infinity. We similarly define $\Lambda^-$ as the intersection of $\mathcal{I}^-$ with the union of all null geodesics with future pointing tangent vector asymptoting to $k^a$ at past null infinity. In this paper we will restrict attention to a quasi-Cartesian frame (Definition \ref{defn:quasicart}) and choose co-ordinates $(x_0,x_1,...,x_D)$ such that $k^a$ has components $k^\mu=(1,0,...,0,1)$ in this frame.  

The key lemma in \cite{PSW} is Lemma III.2.1. This gives an estimate of the \textit{time of flight} along a curve consisting of the union of two null geodesics joined at a point\footnote{Note that our choice of co-ordinates such that $k^\mu=(1,0,...,0,1)$ ensures our curve must pass through such a point.} with $x_D=0$ and with endpoints on $\Lambda^\pm$. The time of flight is defined to be the retarded time of arrival on $\Lambda^+$ minus the advanced time of departure from $\Lambda^-$ (where these are defined using the Hamilton-Jacobi functions describing null geodesics with endpoints on $\mathcal{I}^+$ and $\mathcal{I}^-$ respectively). It is argued that this quantity behaves asymptotically like
\begin{equation}\label{eqn:timeofflightestimate}
    4P\cdot k\log(b/R)
\end{equation}
where $P^a$ is the ADM 4-momentum of the spacetime \cite{ADM}, $b$ is the value of $r:=\sqrt{\sum\limits_{i=1}^{D}x_i^2}$ (defined using quasi-Cartesian co-ordinates) at the point where the null geodesics join, and $R$ is a positive constant. 

Suppose $P\cdot k>0$, i.e. the 4-momentum of the spacetime is not future causal\footnote{We are using the ``mostly plus'' signature (-,+,...,+).}. The aim is to show that there must then exist a fastest causal curve from $\Lambda^-$ to $\Lambda^+$ which enters the interior of the spacetime. If this is the case then this curve must lie on the boundary of the causal future of some point $p\in\Lambda^-$ and hence must be a null geodesic (\cite{Wald} Corollary after Theorem 8.1.2) without conjugate points (\cite{HawkingEllis} Proposition 4.5.12).

The construction begins by finding points $p\in\Lambda^-$, $q\in\Lambda^+$ such that there is no causal curve which departs $\Lambda^-$ later than $p$ and arrives at $\Lambda^+$ earlier than $q$. These will be the endpoints of the fastest causal curve, $\gamma$, to be constructed. Consider a sequence of causal curves, $(\gamma_i)_{i=0}^\infty$, with endpoints on $\Lambda^\pm$ which tend towards $p$ and $q$ and with $\gamma_i$ faster than $\gamma_j$ for $i>j$. The curve $\gamma$ is defined to be the limit of this sequence of causal curves, where we use the fact that, in a globally hyperbolic set, the space of causal curves between two compact sets is compact (\cite{SorkinWoolgar} Theorem 23). 

It remains to check that $\gamma$ does in fact enter the interior of the spacetime and hence define a null line.  The curves $(\gamma_i)_{i=0}^\infty$ can be modified (possibly making them faster) so that they consist of two null geodesics joined at a point with $x_D=0$, $r=b_i$. This means that the estimate (\ref{eqn:timeofflightestimate}) now applies. If we were to have $b_i\longrightarrow\infty$ as $i\longrightarrow\infty$ then this estimate tells us that the time of flight along $\gamma_i$ would also diverge to $+\infty$ as $i\longrightarrow\infty$. In particular, the sequence ($\gamma_i)^\infty_{i=0}$ would eventually become slower than $\gamma_0$. This contradicts the definition of the sequence, so we conclude that $b_i$ must not diverge along the sequence and hence, possibly restricting to a subsequence, all of the $\gamma_i$ must enter the compact set $\mathcal{K}:=\left(J^+(p_0)\cap J^-(q_0)\right)\setminus \mathcal{U}_R$, where $\mathcal{U}_R:=\{x\in J^+(p_0)\cap J^-(q_0):r(x)>R\}$ (see Section \ref{Definitions and Assumptions} for definitions). Once again, using the compactness result of \cite{SorkinWoolgar}, we conclude that $\gamma$ must also enter this set, and hence must enter the interior of the spacetime. We therefore conclude that $\gamma$ is a null line. 

This argument does not generalise to higher dimensions because the time of flight along curves restricted to arbitrarily large values of $r$ no longer diverges. Instead, the time of flight estimate (\ref{eqn:timeofflightestimate}) is replaced by Lemma \ref{lemma:timeofflight}. This lemma says that the time of flight along a curve from $\Lambda^-$ to $\Lambda^+$ which consists of two null geodesics tends to 0 as we let $R\longrightarrow\infty$, where $R$ is such that $r>R$ along the curve.

However, Lemma \ref{lemma:timeofflight} combined with ideas from \cite{PCMD} turns out to be sufficient to construct a null line for negative mass spacetimes in higher dimensions (with some slightly modified assumptions). In particular, we generalise a result of \cite{PCMD} to show that, by a comparison argument involving a Minkowski metric defined on a neighbourhood of conformal infinity, the presence of negative mass allows us to construct a Minkowski-null curve from $\Lambda^-$ to $\Lambda^+$ which is timelike with respect to the physical metric\footnote{This curve will also be timelike with respect to the compactified metric $\tilde{g}$ since conformal transformations preserve the causal structure. We will use this fact throughout when switching between spacetimes with metrics related by a conformal transformation.} $g$. To do this, we show that for negative mass spacetimes, the Minkowski null cones at sufficiently large $r$ are contained inside the $g$-null cones. Defining retarded and advanced time co-ordinates as in equation (\ref{eqn:retadv}), it is a straightforward calculation to show that the time of flight along a Minkowski null geodesic is exactly zero. So, if we choose a $\eta$-null geodesic restricted to sufficiently large values of $r$, then this curve must be timelike with respect to the metric $g$. Then since the timelike future of any point is an open set, it must be possible to modify this curve slightly to obtain a $g$-timelike curve, $\gamma_0$, between $\Lambda^-$ and $\Lambda^+$ which has time of flight strictly less than zero. This allows us to use a similar construction as was used in $3+1$ dimensions based on a sequence of faster and faster causal curves $(\gamma)_{i=0}^\infty$ from $\Lambda^-$ to $\Lambda^+$. If $b_i\longrightarrow\infty$ along this sequence, then by Lemma \ref{lemma:timeofflight} the time of flight will tend to 0 and in particular will eventually become larger than the time of flight of $\gamma_0$. As in 3+1 dimensions, this allows us to conclude that a fastest causal curve from $\Lambda^-$ to $\Lambda^+$ does exist. 

\section{Definitions and Assumptions}\label{Definitions and Assumptions}
In order to carry out the comparison with Minkowski spacetime mentioned in the previous section, it will be necessary to impose stronger conditions than those used in \cite{PSW}. These conditions will be more similar to the ones used in \cite{poorman} and \cite{poorman2}. In this section we outline the various assumptions made.
\begin{definition}\label{defn:spacetime}
A \textit{spacetime} $(M,g)$ is a connected manifold, $M$, of dimension $D+1$ ($D\geq3$) equipped with a $C^{1,1}$ Lorentzian metric $g$ of signature $(D,1)$.
\end{definition}
For the purposes of Theorem \ref{thm:nullline}, requiring the metric to be $C^{1,1}$ will be sufficient. In order to obtain a focusing result in Section \ref{A Focusing Theorem in Higher Dimensions}, it may be necessary to make stronger assumptions. For example, in Theorem \ref{thm:Borde} we assume that the quantity $R_{ab}T^aT^b$ is finite and continuous, where $T^a$ is tangent to a null geodesic. To ensure this, it would be sufficient to assume that the metric is $C^2$.
\begin{definition}\label{defn:quasicart}
A spacetime $(M,g)$ admits \textit{quasi-Cartesian co-ordinates} if there are co-ordinates, defined on some subset of $M$ diffeomorphic to $\mathbbm{R}\times\mathbbm{R}^D\setminus B$ (where $B$ denotes a closed ball in $\mathbbm{R}^D$), with respect to which the components of the metric, $g$, take the form
\begin{equation}
    g_{\mu\nu}=\eta_{\mu\nu}+h_{\mu\nu}
\end{equation}
For our purposes it will be sufficient to assume that
\begin{equation}
    \begin{split}
        h_{\mu\nu}&=O\left(r^{-\alpha}\right)\\
        \partial_\rho h_{\mu\nu}&=O\left(r^{-(1+\alpha)}\right)
    \end{split}
\end{equation}
for some $\alpha>1$.
\end{definition}
Note that Schwarzschild spacetime in $3+1$ dimensions does not satisfy these conditions since in this case we have $h_{\mu\nu}=O(r^{-1})$ and $\partial_\rho h_{\mu\nu}=O(r^{-2})$. As a result, this spacetime is not covered by the results of this paper. However, the conditions stated above are satisfied by Schwarzschild in higher dimensions. 


Requiring that a spacetime admits quasi-Cartesian co-ordinates will allow us to prove Lemma \ref{lemma:timeofflight}, however in order to construct a null line we will need to consider a more restrictive class of metrics. Recall that the $D+1$ dimensional Schwarzschild metric with ADM mass $m_{ADM}\in\mathbbm{R}$, which we denote $g_m$, has line element
\begin{equation}\label{eqn:Schwarzschild}
    ds_m^2=-\left(1-\frac{2m}{r^{D-2}}\right)dt^2+\frac{dr^2}{1-\frac{2m}{r^{D-2}}}+r^2d\omega^2_{D-1}
\end{equation}
where $d\omega^2_{D-1}$ is the round line element on the unit $(D-1)$-sphere, $S^{D-1}$, and the \textit{mass parameter}, $m$, is related to $m_{ADM}$ by\footnote{We work in units in which $c=G=1$.}
\begin{equation}\label{eqn:mass}
    m_{ADM}=\frac{(D-1)Area(S^{D-1})}{8\pi}m
\end{equation}
\begin{definition}\textbf{(Following \cite{poorman})} \label{defn:asymptoticallyschwarzschildean}
For $m\in \mathbbm{R}$, we say that a metric $g$ on $\mathbbm{R}\times\left(\mathbbm{R}^D\setminus B\right)$, where $B$ is a ball of radius $R$ with $R^{D-2}>2m$, is \textit{uniformly Schwarzschildean} if, in the co-ordinates of (\ref{eqn:Schwarzschild}) (or equivalently in the co-ordinates of equation (1) of \cite{poorman}):
\begin{equation}
\begin{split}
        g-g_m&=o\left(|m|r^{-(D-2)}\right)\\
    \partial_i(g-g_m)_{jk}&=o\left(|m|r^{-(D-1)}\right)
\end{split}
\end{equation}
As in \cite{poorman} we will abuse notation and allow $m=0$ in this definition, by which we mean the metric is flat for $r>R$, for some $R\in\mathbbm{R}_{\geq0}$.\footnote{This case will not be important for us since we will be unable to comment on the $m=0$ case.}
\end{definition}
Note that the conditions imposed on the spacetime are less restrictive than the ones used in \cite{poorman2}, where the spacetime is assumed to be \textit{strongly uniformly Schwarzschildean}. 

Throughout this paper we will refer to the following sets:
\begin{equation}
    \begin{split}
        J^+(p)&=\{q\in\tilde{M}|\exists\text{ a smooth future-directed causal curve from $p$ to $q$}\}\\
        J^-(p)&=\{q\in\tilde{M}|p\in J^+(q)\}\\
        I^+(p)&=\{q\in\tilde{M}|\exists\text{ a smooth future-directed timelike curve from $p$ to $q$}\}\\
        I^-(p)&=\{q\in\tilde{M}|p\in I^+(q)\}\\
    \end{split}
\end{equation}
For non-compact spacetimes, $(M,g)$, we can define the same sets as subsets of $M$ rather than as subsets of $\tilde{M}$. As in \cite{PSW}, we define single points to be curves of zero length, so $p\in J^\pm(p)$ but $p\notin I^\pm(p)$.

In Theorem \ref{thm:nullline} we will require that $\mathcal{D}\cup\mathcal{I}$ be globally hyperbolic as a subset of $\tilde{M}$, where $\mathcal{D}$ denotes the \textit{domain of outer communications} $\mathcal{D}=I^-(\mathcal{I}^+)\cap I^+(\mathcal{I}^-)$ and $\mathcal{I}=\mathcal{I}^+\cup\mathcal{I}^-$. By this we mean that $\mathcal{D}\cup\mathcal{I}$ is strongly causal and contains $J^+(p)\cap J^-(q)$ as a compact subset for each $p,q\in\mathcal{D}\cup\mathcal{I}$ (\cite{HawkingEllis} Section 6.6).

Following \cite{HawkingEllis}, we make the following definition:
\begin{definition}\label{defn:asymptotically empty and simple}
A spacetime $(M,g)$ is 
\textit{asymptotically empty and simple} if there is a strongly causal spacetime $(\tilde{M},\tilde{g})$ and an embedding $\theta:M\longrightarrow\tilde{M}$ which embeds $M$ as a manifold with smooth boundary $\partial M$ in $\tilde{M}$, such that 
\begin{enumerate}
\item there is a smooth function $\Omega$ on $\tilde{M}$ such that on $\theta(M)$, $\Omega$ is positive and $\Omega^2g=\theta_*(\tilde{g});$
\item $\Omega=0$ and $d\Omega\neq0$ on $\partial M$;
\item $R_{ab}=0$ in an open neighbourhood of $\mathcal{I}$ in $\tilde{M}$; and
\item every null geodesic in $M$ acquires a future and past endpoint on $\mathcal{I}$.
\end{enumerate}
\end{definition}

If $(M,g)$ is an asymptotically empty and simple spacetime, then $\partial M $ is a null surface and can be split into two parts \cite{HawkingEllis}: past null infinity, denoted $\mathcal{I}^-$, where null geodesics have their past endpoints, and future null infinity, denoted $\mathcal{I}^+$, where null geodesics have their future endpoints.

It is common to also label the following points, which lie in the topological boundary of $\tilde{M}$:
\begin{itemize}
       \item future timelike infinity, $i^+$: the point consisting of the future endpoints of timelike geodesics;
    \item past timelike infinity, $i^-$: the point consisting of the past endpoints of timelike geodesics;
    \item and spatial infinity, $i^0$: the point consisting of the endpoints of spacelike geodesics.
\end{itemize}

The final condition in Definition \ref{defn:asymptotically empty and simple} is extremely restrictive, since it rules out spacetimes containing black hole regions. We will instead consider spacetimes which are \textit{weakly asymptotically empty and simple} \cite{HawkingEllis}.
\begin{definition}\label{defn:weakly asymptotically empty and simple}
A spacetime $(M,g)$ is \textit{weakly asymptotically empty and simple} if there is an asymptotically empty and simple spacetime $(M',g')$ and a neighbourhood $U'$ of $\partial M '$ in the corresponding $\tilde{M}'$ such that $U'\cap M'$ is isometric to a subset of $M$.
\end{definition}

For a weakly asymptotically empty and simple spacetime, $(M,g)$, we define the conformal boundary at infinity, denoted $\mathcal{I}$, to be the points in $\partial M $ which are identified with $\partial M '$ by the isometry in Definition \ref{defn:weakly asymptotically empty and simple}. This can then be split into two parts, $\mathcal{I}^+$ and $\mathcal{I}^-$, as for an asymptotically empty and simple spacetime, again using the isometry in Definition \ref{defn:weakly asymptotically empty and simple}. The points $i^+$, $i^-$ and $i^0$ in the topological boundary of $\tilde{M}$ can also be labelled similarly.


As in [14], we assume $\tilde{M}$ extends slightly past conformal infinity so that it is indeed a manifold. This is required in order to satisfy the conditions of Theorem 23 in \cite{SorkinWoolgar} which is used in the proof of Theorem \ref{thm:nullline}.


We will require the following results regarding the completeness of $\mathcal{I}^\pm$.
\begin{prop}(\cite[Proposition 6.9.4]{HawkingEllis})
In a ($D+1$)-dimensional asymptotically simple and empty spacetime $(M,g)$, $\mathcal{I}^+$ and $\mathcal{I}^-$ are topologically $\mathbbm{R}\times S^{D-1}$ and $M$ is $\mathbbm{R}^{D+1}$.
\end{prop}
This tells us that in an asymptotically simple and empty spacetime, $\mathcal{I}^\pm$ are the same as in Minkowski spacetime of the same dimension. The following corollary follows immediately from this proposition and from definition \ref{defn:weakly asymptotically empty and simple}.
\begin{cor}
In a ($D+1$)-dimensional weakly asymptotically simple and empty spacetime $(M,g)$, $\mathcal{I}^+$ and $\mathcal{I}^-$ are topologically $\mathbbm{R}\times S^{D-1}$.
\end{cor}

\section{Time of Flight Estimate in Higher Dimensions}\label{time of flight Estimate in Higher Dimensions}
In this section we will derive a higher dimensional analogue of \cite[Lemma III.2.1]{PSW} which gives an estimate for the time of flight of causal curves near infinity with endpoints on $\Lambda^\pm$. In \cite{PSW} it was found that the time of flight diverged logarithmically as we considered curves restricted to increasingly large values of $r$. We will show that in higher dimensions, the time of flight instead tends to $0$. The absence of a divergence is the reason the 3+1 dimensional argument given in \cite{PSW} could not be generalised to higher dimensions. 

We will show that higher dimensional spacetimes admitting quasi-Cartesian co-ordinates can be compactified using the same procedure (and same retarded and advanced time co-ordinates) as Minkowski spacetime. As a result, the time of flight along curves with endpoints on $\mathcal{I}^\pm$ can be calculated as the difference between the Minkowski retarded and advanced time co-ordinates, $u=t-r$ and $v=t+r$, evaluated at future and past null infinity respectively. This method avoids the need to consider Hamilton-Jacobi functions $S^\pm$, as is done in \cite{poorman2} and \cite{PSW}.

We begin by recalling how Minkowski spacetime can be compactified \cite[Section 3]{PCMD}. We define retarded and advanced time co-ordinates
\begin{equation}\label{eqn:retadv}
\begin{split}
     u:=t-r,\quad    v:=t+r
\end{split}
\end{equation}
and then compactify the metric by defining
\begin{equation}\label{eqn:retardedadvanced}
u=\tan P, \quad
v=\tan Q
\end{equation}
Finally, we define new co-ordinates 
\begin{equation}\label{eqn:Tchi}
 T=Q+P\in(-\pi,\pi), \quad
        \chi=Q-P\in[0,\pi)
    \end{equation}
which we can think of as ``time" and ``radial" co-ordinates respectively in the compactified spacetime.

We then consider the conformally related metric, $\tilde{g}$, given by
\begin{equation}
    \tilde{g}=\Omega^2g=\left(2\cos P\cos Q\right)^2g
\end{equation}
with corresponding line element
\begin{equation}\label{eqn:compactMink}
    \Tilde{ds}^2=-dT^2+d\chi^2+\sin^2\chi d\omega_{D-1}^2
\end{equation}
\begin{figure}
    \centering
    \includegraphics[scale=0.3]{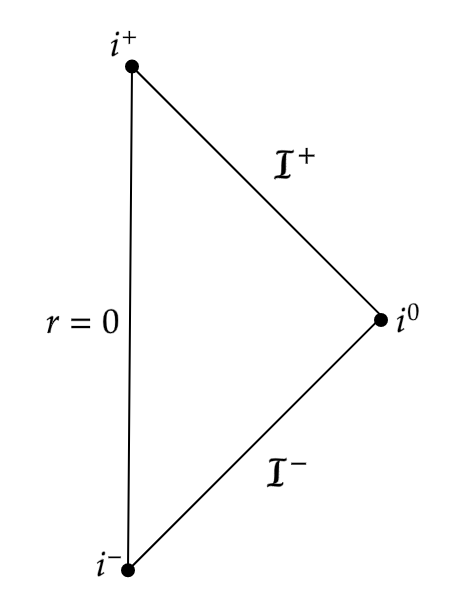}
    \caption{Figure showing the Penrose diagram for Minkowski spacetime.}
    \label{fig:compactminkowski}
\end{figure}

The conformal boundary at infinity can be split into two parts, as described in section \ref{Definitions and Assumptions}, as follows:
\begin{equation}\label{eqn:conformalboundarysplit}
    \begin{split}
        \mathcal{I}^+&:=\{u\in\mathbbm{R}, v=\infty\}\\
        \mathcal{I}^-&:=\{u=-\infty, v\in\mathbbm{R}\}
    \end{split}
\end{equation}
The points $i^+$, $i^-$ and $i^0$ are given by
\begin{equation}
    \begin{split}
                i^+&:=\{u=\infty, v=\infty\}\\
        i^-&:=\{u=-\infty, v=-\infty\}\\
        i^0&:=\{u=-\infty, v=\infty\}
    \end{split}
\end{equation}
This is illustrated in Figure \ref{fig:compactminkowski}.

If this procedure is carried out for Schwarzschild spacetime of mass $m$ in 3+1 dimensions, the result is a spacetime where null geodesics which do not cross an event horizon have endpoints at $i^\pm$ if $m>0$ (they are infinitely delayed), or are located entirely at $i^0$ if $m<0$ (they are infinitely advanced). This violates condition 4 of Definition \ref{defn:asymptotically empty and simple}. Instead, the standard procedure for the compactification of Schwarzschild \cite{PCMD} involves re-scaling by a factor of $\frac{1}{V(r)}$ and defining retarded and advanced time co-ordinates by 
\begin{equation}\label{eqn:Schwarzschildnulladv}
    \begin{split}
        u_s&=t-r_*\\
        v_s&=t+r_*
    \end{split}
\end{equation}
where 
\begin{equation}
\begin{split}
\frac{dr_*}{dr}&=\frac{1}{V(r)}\\
       \implies r_*&=\begin{cases}r+2m\log(r/2m-1)&\text{ if }D=3\\
    r+O(r^{3-D})&\text{ if }D\geq4
    \end{cases}
\end{split}
\end{equation}

In 3+1 dimensions, $r_*$ diverges from $r$ logarithmically. This is the reason we do not obtain a good compactification using Minkowski retarded and advanced time co-ordinates (in the sense that null geodesics escaping to the asymptotic region do not have endpoints on the surfaces $\mathcal{I}^\pm$ as defined in (\ref{eqn:conformalboundarysplit})). We see that in higher dimensions, $u_s$ and $v_s$ agree with the Minkowski $u$ and $v$ at $r=\infty$. Consequently, we could alternatively have used these Minkowski retarded and advanced time co-ordinates to compactify Schwarzschild in higher dimensions, since the structure of the conformal boundary at infinity would be the same in both cases. The following lemma (similar to \cite[Proposition B.1]{invariantmass}) shows that this is a general feature of higher dimensional spacetimes admitting quasi-Cartesian co-ordinates.

\begin{lemma}\label{lemma:compactification}
Let $(M,g)$ be a spacetime in $D+1$ dimensions ($D\geq4$) which admits quasi-Cartesian co-ordinates. Let $\gamma$ be a future endless $g$-null geodesic segment with co-ordinates $x^\mu(s)$, where $s\geq0$ is an affine parameter (increasing to the future). Suppose $r\longrightarrow\infty$ as $s\longrightarrow\infty$. Then
$\dot{x}^\mu(s)$ tends to a finite limit as $s\longrightarrow\infty$, denoted $\dot{x}^\mu_\infty$, (where $\dot{}$ denotes differentiation with respect to $s$). Moreover, if $\gamma$ lies entirely in the the region $r>R$ then there exists some constant $C$ such that
\begin{equation}\label{eqn:errorbound}
    \left|x^\mu(s)-\dot{x}^\mu_\infty 
    s-x^\mu(0)\right|\leq \frac{C}{R^{\alpha-1}}
\end{equation}
for any $s\geq0$ and any index $\mu$. In particular, Minkowski retarded time $u:=t-r$ tends to a finite limit as $s\longrightarrow\infty$.

Similarly, if $\gamma$ is instead past endless and the affine parameter $s$ is chosen so that $s\leq0$ along $\gamma$, then $\dot{x}^\mu(s)$ tends to a finite limit, denoted $\dot{x}^\mu_{-\infty}$, as $s\longrightarrow{-\infty}$. Moreover, if $\gamma$ lies entirely in the region $r>R$ then there exists some constant $C'$ such that 
\begin{equation}\label{eqn:errorboundpast}
    \left|x^\mu(s)-\dot{x}^\mu_{-\infty}
    s-x^\mu(0)\right|\leq \frac{C'}{R^{\alpha-1}}
\end{equation}
holds for any $s\leq0$ and any index $\mu$. In particular, Minkowski advanced time $v:=t+r$ tends to a finite limit as $s\longrightarrow-\infty$.
\end{lemma}
 



\textbf{Proof}: Let $R$ be such that $r\geq R$ for all $s\geq0$. Since $r\longrightarrow\infty$ as $s\longrightarrow\infty$, by shifting the origin of $s$ we are free to make $R$ arbitrarily large and enforce $\frac{dr^2}{ds}|_{s=0}\geq0$. Next, re-scale $s$ so that $\frac{dx^i}{ds}\frac{dx^i}{ds}\vert_{s=0}=1$.  Let $s_1>0$ be maximal such that $\frac{3}{4}<\frac{dx^i}{ds}\frac{dx^i}{ds}<\frac{5}{4}$ for all $0\leq s<s_1$. From the geodesic equations, we have 
\begin{equation}\label{eqn:doublederivrsquared}
    \frac{d^2r^2}{ds^2}=2\left(\frac{dx^i}{ds}\frac{dx^i}{ds}+x^i\Gamma^i_{\mu\nu}\frac{dx^\mu}{ds}\frac{dx^\nu}{ds}\right).
\end{equation}
Since $\gamma$ is null, for $R$ sufficiently large and $0\leq s<s_1$ we have $|dt/ds|<2$ and hence $|dx^\mu/ds|$ is bounded for $\mu=0,1,...,D$.  Since $(M,g)$ admits quasi-Cartesian co-ordinates, we also have
\begin{equation}
    \left|\Gamma^\mu_{\nu\rho}\right|\leq C_1r^{-\alpha-1}
\end{equation}
for some constant $C_1$.

Substituting this into equation (\ref{eqn:doublederivrsquared}), we have
\begin{equation}
    \frac{d^2r^2}{ds^2}\geq 2\times\left(\frac{3}{4}-C_2r^{-\alpha}\right)
\end{equation}
for some constant $C_2$. Hence for $0< s<s_1$ and $r\geq R$ (increasing $R$ if necessary), we have 
\begin{equation}
\begin{split}
    \frac{d^2r^2}{ds^2}&>1\\
    \implies r^2(s)&\geq r^2(0)+s\left.\frac{dr^2}{ds}\right|_{s=0}+\frac{s^2}{2}\\
    &\geq R^2+\frac{s^2}{2}\\
    &> C_3(R+s)^2
\end{split}
\end{equation}
for some constant $C_3>0$. To derive the final inequality above, we note that if $C_3$ is chosen to be sufficiently small, then this inequality holds for $s=0$ and the equation
\begin{equation}
   \left(\frac{1}{2}-C_3\right)s^2-2C_3Rs+(1-C_3)R^2=0,
\end{equation}
viewed as a quadratic in $s$, has no real solutions.

From this it follows that, for $0\leq s<s_1$:
\begin{equation}\label{eqn:derivativebound}
    \begin{split}
        \left|\int^{s}_0\frac{d^2x^\mu}{ds'^2}ds'\right|&\leq\int^{s}_0\left|\frac{d^2x^\mu}{ds'^2}\right|ds'\\
        \implies \left|\frac{dx^\mu}{ds'}(s)-\frac{dx^\mu}{ds'}(0)\right|&\leq\int_0^{s}\left|\Gamma^\mu_{\nu\rho}\frac{dx^\nu}{ds'}\frac{dx^\rho}{ds'}\right|ds'\\
        &\leq C_4\int^{s}_0r^{-\alpha-1}ds'\\
        &< C_5\int^{s}_0(R+s')^{-\alpha-1}ds'\\
        &\leq C_6R^{-\alpha}\\
        \implies \left|\sum_{i=1}^d\frac{dx^i}{ds}\frac{dx^i}{ds}(s)-1\right|&\leq C_7R^{-\alpha}
    \end{split}
\end{equation}
where $C_4, C_5, C_6, C_7>0$ are constants. To obtain the final inequality above, we have used the fact that if $\textbf{x},\textbf{y}\in(\mathbbm{R}^d,\delta)$, where $\delta$ denotes the Euclidean metric, with $|\textbf{x}|\leq K$ and $|\textbf{y}|=1$, then we have
\begin{equation}
\begin{split}
    \left|\textbf{x}^2-\textbf{y}^2\right|&=\left|(\textbf{x}+\textbf{y})\cdot(\textbf{x}-\textbf{y})\right|\\
    &\leq \left|\textbf{x}+\textbf{y}\right|\left|\textbf{x}-\textbf{y}\right|\\
    &\leq (1+K)\left|\textbf{x}-\textbf{y}\right|.
    \end{split}
\end{equation}

We conclude that by choosing $R$ sufficiently large, we can take $s_1=\infty$.

If, rather than integrating from 0 to $s$ in (\ref{eqn:derivativebound}), we instead integrate between $s_2$ and $s_3>s_2$, we find that 
\begin{equation}\label{eqn:derivativebound2}
    \begin{split}
        \left|\frac{dx^\mu}{ds}(s_3)-\frac{dx^\mu}{ds}(s_2)\right|
        &\leq C_6(R+s_2)^{-\alpha} \text{ for all }s_3>s_2
    \end{split}
\end{equation}
The right hand side of this inequality tends to 0 as $s_2\longrightarrow\infty$, so we conclude that $\dot{x}^\mu(s)$ tends to a finite limit as $s\longrightarrow\infty$. We denote this limit by $\dot{x}^\mu_\infty$.

Then for any $s\geq0$ and any index $\mu$, we have:
\begin{equation}\label{eqn:errorbound2}
\begin{split}
    \left|\int_0^{s}\frac{dx^\mu}{ds'}-\dot{x}^\mu_\infty ds'\right|&\leq \int_0^{s}\left|\frac{dx^\mu}{ds'}-\dot{x}^\mu_\infty\right|ds'\\
    \implies\left|x^\mu(s)-\dot{x}^\mu_\infty 
    s-x^\mu(0)\right|&\leq\int_0^{s} \frac{C_6}{(R+s')^\alpha}ds'\\
&\leq \frac{C}{R^{\alpha-1}}
\end{split}
\end{equation}
where $C=\frac{C_6}{\alpha-1}>0$ is a constant. 

Asymptotic flatness implies that $\eta_{\mu\nu}\dot{x}_\infty^\mu\dot{x}_\infty^\nu=0$ and hence that the curve $x_{Mink}^\mu(s):=x(0)^\mu+\dot{x}^\mu_\infty s$ defines a Minkowski null geodesic along which $u:=t-r$ tends to a finite value as $s\longrightarrow\infty$. From (\ref{eqn:errorbound2}) we have
\begin{equation}\label{eqn:minkcomparison}
    \left|x^\mu(s)-x^\mu_{Mink}(s)\right|\leq\frac{C}{R^{\alpha-1}}
\end{equation}
so we conclude that $u$ must also tend to a finite limit along $\gamma$ as $s\longrightarrow\infty$.

If $\gamma$ is instead a past endless null geodesic segment then similar arguments can be used to show that $\dot{x}^\mu(s)$ tends to a finite limit as $s\longrightarrow-\infty$ and to derive (\ref{eqn:errorboundpast}). One can then deduce that $v:=t+r$ tends to a finite limit as $s\longrightarrow-\infty$. \qedsymbol



Lemma \ref{lemma:compactification} tells us that if we compactify such a spacetime using the same procedure as used for Minkowski (including the same retarded and advanced time co-ordinates), then null geodesics escaping to the asymptotic region in the infinite future (respectively past) will have endpoints on the null surface $\mathcal{I}^+$ (respectively $\mathcal{I}^-$) as defined in (\ref{eqn:conformalboundarysplit}). This can be summarised in the following corollary.

\begin{cor}
Let $(M,g)$ be a spacetime in $D+1$ dimensions ($D\geq4$) admitting quasi-Cartesian co-ordinates. Then $(M,g)$ is weakly asymptotically simple and empty and furthermore the compactification map $\theta$ in Definition \ref{defn:asymptotically empty and simple} can be taken to be the same as for Minkowski spacetime.
\end{cor}

The above results allow us to define the time of flight for spacetimes admitting quasi-Cartesian co-ordinates as follows:
\begin{definition}\label{defn:timeofflight}
Let $(M,g)$ be a spacetime in $D+1$ dimensions ($D\geq4$) admitting quasi-Cartesian co-ordinates. We define the (possibly negative) time of flight along an endless curve as $u_\infty-v_\infty$, where $u_\infty$ denotes the value of $u=t-r$ at the future endpoint of the curve and $v_\infty$ denotes the value of $v=t+r$ at its past endpoint.
\end{definition}
Lemma \ref{lemma:compactification} tells us that the time of flight is finite along null geodesics which escape to the asymptotic region in both the future and the past (so in particular do not enter black or white holes). In \cite{PSW}, the time of flight is defined to be $S^+-S^-$, where the Hamilton-Jacobi functions $S^+$ and $S^-$ are finite on future and past null infinity respectively. According to \cite{poorman2}, the proof of existence of the optical functions $S^\pm$ is the missing step of the argument in \cite{PSW}. The above definition means that it is not necessary to define such functions here.


We can now prove the following time of flight estimate for spacetimes admitting quasi-Cartesian co-ordinates:
\begin{lemma}(\textbf{Time of Flight Estimate}) \label{lemma:timeofflight}
Let $(M,g)$ be a spacetime in $D+1$ dimensions ($D\geq4)$ which admits quasi-Cartesian co-ordinates and let $\gamma$ be a causal curve which connects $\Lambda^-$ to $\Lambda^+$ (see Section \ref{Introduction} for definitions) and is comprised of two null geodesic segments. Suppose $\gamma$ lies entirely in the region $r>R$. Then, for $R$ sufficiently large, the time of flight along $\gamma$ satisfies 
\begin{equation}
    |u_\infty-v_\infty|\leq\frac{A}{R^{\alpha-1}}
\end{equation}
for some constant $A$.
\end{lemma}

\textbf{Proof}: Let $p_*$ denote the point at which the two null geodesic segments are joined. Let $\gamma_+$ denote the null geodesic from $p_*$ to $\Lambda^+$ and let $\gamma_-$ denote the null geodesic from $\Lambda^-$ to $p_*$. Let $\gamma_{\eta}$ denote the curve through $p_*$ with co-ordinates
\begin{equation}
    x_{\eta}^\mu(s)=x_{p_*}^\mu+ sk^\mu
\end{equation}
where $k^\mu=(1,0,...,0,1)$ is the Minkowski-null direction used to define $\Lambda^\pm$.

\begin{figure}
    \centering
    \includegraphics[scale=0.3]{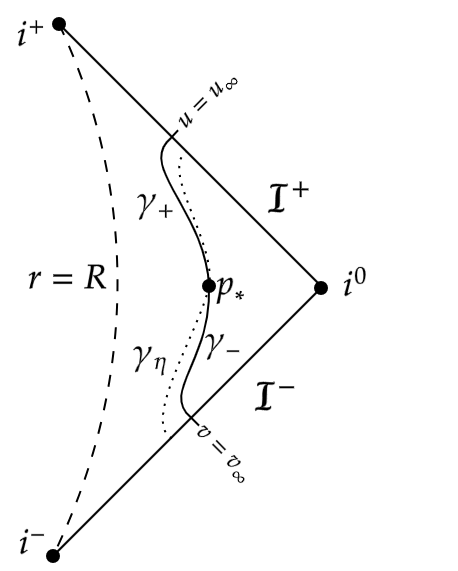}
    \caption{Lemma \ref{lemma:timeofflight} shows that if $\gamma_-$ is a $g$-null geodesic from $\Lambda^-$ to $p_*$ and $\gamma_+$ is a $g$-null geodesic from $p_*$ to $\Lambda^+$, with both restricted to the region $r>R$, then the time of flight along $\gamma_-\cup\gamma_+$ approaches 0 as $R\longrightarrow\infty$. The lemma is proved by comparing this time of flight to the time of flight along a Minkowski null geodesic, $\gamma_\eta$, through $p_*$, along which the time of flight is 0.}
    \label{fig:timeofflightestimate}
\end{figure}

Choose the affine parameter $s$ so that $\frac{dx^i}{ds}\frac{dx^i}{ds}|_{s=0}=1$ (as in the proof of Lemma \ref{lemma:compactification}). We will use inequality (\ref{eqn:minkcomparison}), which tells us that for $R$ sufficiently large, the time of flight along $\gamma$ is close to the time of flight along $\gamma_{\eta}$, which we now calculate. 


Let $u_{\gamma_\eta,\infty}$ denote the $u$ co-ordinate at the future endpoint of $\gamma_\eta$; $b$ denote the impact parameter of $\gamma_\eta$ (the smallest value of $r$ attained along $\gamma_\eta$); and $t_b$ denote the value of the $t$ co-ordinate on $\gamma_\eta$ when $r=b$. Suppose $s=s_b$ when $t=t_b$ and $r=b$. Along $\gamma_\eta$ we have $\dot{t}=1$ and hence 
\begin{equation}
    \begin{split}
    t(s)&=t_b+s-s_b.
        \end{split}
\end{equation}
Furthermore, $x^D=0$ at $r=b$, so we have 
    \begin{equation}
        \begin{split}
    r(s)&=\sqrt{b^2+(s-s_b)^2}\\
        \implies u(s)&:=t(s)-r(s)\\
        &=t_b+s-s_b-\sqrt{b^2+(s-s_b)^2}\\
\implies u_{\gamma_\eta,\infty} &:=\lim_{s\longrightarrow\infty}u(s)\\
&=t_b
\end{split}
    \end{equation}

Similarly, the advanced time co-ordinate at the past endpoint of $\gamma_\eta$ is $v_{\gamma_\eta,\infty}=t_b$. Hence the time of flight along $\gamma_\eta$ is
\begin{equation}
\begin{split}
    u_{\gamma_\eta,\infty}-v_{\gamma_\eta,\infty}&=0
\end{split}
\end{equation}
Consequently, it follows from Lemma \ref{lemma:compactification} that the time of flight along $\gamma$ is
\begin{equation}
\begin{split}
    |u_\infty-v_\infty|&\leq |u_\infty-u_{\gamma_\eta,\infty}|+|u_{\gamma_\eta,\infty}-v_{\gamma_\eta,\infty}|+|v_{\gamma_\eta,\infty}-v_\infty|\\
    &\leq \frac{A}{R^{\alpha-1}}\\
\end{split}
\end{equation}
for some constant $A>0$. \qedsymbol 

Note that the above argument relied on the fact that defining the time of flight along $\gamma$ using Minkowski retarded and advanced time co-ordinates gives a finite result (Lemma \ref{lemma:compactification}). For Schwarzschild in $3+1$ dimensions, this time of flight is $\pm\infty$ for $m\gtrless0$. As a result, although the metric is asymptotically flat, we cannot conclude that the time of flight approaches $0$ if $\gamma$ is restricted to arbitrarily large $r$.  
\section{Constructing a Fastest Causal Curve}\label{Constructing a Fastest Causal Curve}
As discussed in Section \ref{Introduction}, we will begin by constructing a causal curve from $\Lambda^-$ to $\Lambda^+$ which has negative time of flight. To do this we use a comparison argument based on Minkowski spacetime, similar to the ones used in \cite{PCMD}. This argument relates to the compactified spacetimes. 

\begin{lemma}\label{lemma:negativetimeofflight}
Let $(M,g)$ be a uniformly Schwarzschildean spacetime in $D+1$ dimensions ($D\geq4$) with ADM mass $m_{ADM}<0$. Then there exists an endless timelike curve from $\Lambda^-$ to $\Lambda^+$ which has negative time of flight.
\end{lemma}
Note that a uniformly Schwarzschildean spacetime in higher dimensions necessarily admits quasi-Cartesian co-ordinates. This means that the results of the previous section apply. In particular, we define the time of flight along curves with endpoints on $\mathcal{I}^-$ and $\mathcal{I}^+$ using Definition \ref{defn:timeofflight}, with $u=t-r$ and $v=t+r$. 

\textbf{Proof}: The metric is uniformly Schwarzschildean, so the line element can be written in some neighbourhood of conformal infinity as
\begin{equation}\label{eqn:lineeleement}
    \begin{split}
        ds^2=ds^2_{Mink}+\frac{2m}{r^{D-2}}\left(dt^2+dr^2\right)+o\left(r^{-(D-2)}\right)
    \end{split}
\end{equation}
where by $o\left(r^{-(D-2)}\right)$ we mean that this term is equal to $g'_{\mu\nu}dx^\mu dx^\nu$ for some $g'_{\mu\nu}=o\left(r^{-(D-2)}\right)$.

Since these co-ordinates are defined on some region diffeomorphic to $\mathbbm{R}\times\left(\mathbbm{R}^D\setminus B\right)$, we are able to identify curves in this region with curves in Minkowski spacetime (we simply identify the quasi-Cartesian co-ordinates with some Cartesian co-ordinate system in Minkowski). Furthermore, since the spacetime $(M,g)$ can be compactified to $(\tilde{M},\tilde{g})$ using the same compactified co-ordinates as Minkowski, we can also identify curves in a neighbourhood of $\mathcal{I}$ in $(\tilde{M},\tilde{g})$ with curves in compactified Minkowski (we simply identify the compactified co-ordinates). 

It is clear from (\ref{eqn:lineeleement}) that for sufficiently large $r$, the term proportional to $m$ will dominate the other non-Minkowski terms. This means that if $m_{ADM}<0$ (or equivalently $m<0$), then there exists some $R_0$ such that 
\begin{equation}\label{eqn:bound}
    \begin{split}
        ds^2<ds^2_{Mink}
    \end{split}
\end{equation}
along any Minkowski causal curve at $r>R_0$.

Consider a Minkowski null geodesic with past endpoint at $v=0$ on $\Lambda^-$, future endpoint at $u=0$ on $\Lambda^+$ and impact parameter $b>R_0$. This curve is restricted to the region where $r> R_0$ and hence inequality (\ref{eqn:bound}) guarantees that, after identifying compactified co-ordinates, it defines a $g$-timelike curve in $(\Tilde{M},\Tilde{g})$.

\begin{figure} 
    \centering
    \includegraphics[scale=0.3]{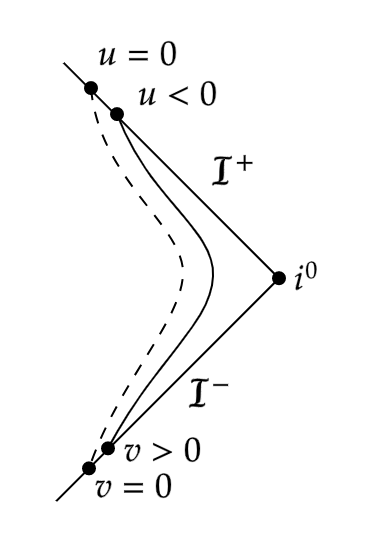}
    \caption{The Minkowski null geodesic between $\Lambda^-$ and $\Lambda^+$ with endpoints at $u=0$ and $v=0$ respectively and impact parameter $b$ is shown as a dotted line. If the metric $g$ is uniformly Schwarzschildean with negative mass and $b$ is sufficiently large, then this path can be modified so that it still connects $\Lambda^-$ to $\Lambda^+$ but is $g$-timelike and has strictly negative time of flight. This modification is shown as an unbroken line.}
    \label{fig:negativeToF}
\end{figure}

Now since $I^\pm(p)$ are open sets for any point $p\in \tilde{M}$, we must be able to modify this curve slightly so that it remains $g$--timelike but has past endpoint at $v>0$ on $\Lambda^-$ and future endpoint at $u<0$ on $\Lambda^+$. This ensures that the curve has negative time of flight (see Figure \ref{fig:negativeToF}).  \qedsymbol

This curve will be the starting point for our sequence of faster and faster causal curves used in the proof of Theorem \ref{thm:nullline} to construct a fastest causal curve from $\Lambda^-$ to $\Lambda^+$. We will also use the time of flight estimate, Lemma \ref{lemma:timeofflight}, in this construction. However in order to do so it is necessary to prove the following lemma.
\begin{lemma}\label{lemma:confinetolargeR} Let $(M,g)$ be a $D+1$ dimensional uniformly Schwarzschildean spacetime ($D\geq4$) and let $R>0$. Let $p_*$ be a point with $t=0$ and let $q\in\mathcal{I}^+$. Then there exists a constant $R'$ such that if $p_*$ has $r$ co-ordinate $r_{p_*}>R'$, then any $g$-null geodesic from $p_*$ to $q$ is confined to the region $r>R$.
\end{lemma}
\begin{figure}
    \centering
    \includegraphics[scale=0.3]{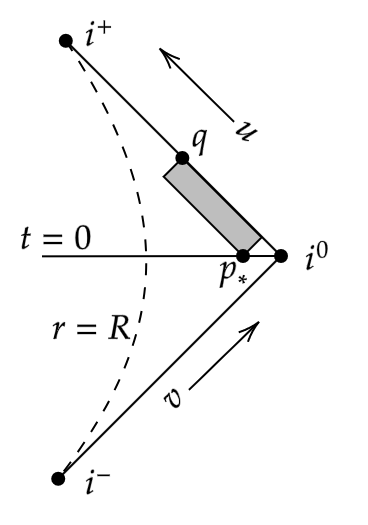}
    \caption{The causal diamond $J^+(p_*)\cap J^-(q)$ (taken with respect to the Minkowski metric) is contained in the shaded region bound by curves of constant $u$ or $v$. These curves are straight lines at $45^\circ$ in the Penrose diagram above. Given some $R>0$, if the point $p_*$ lies sufficiently close to $i^0$ (i.e. it has sufficiently large $r$ co-ordinate), then causal curves from $p_*$ to $q$ remain close to $\mathcal{I}^+$ on this diagram and hence cannot enter the region $r\leq R$.}
    \label{fig:causaldiamond}
\end{figure}
The Minkowski metric in $D+1$ dimensions is
\begin{equation}
    ds^2=-dt^2+dr^2+r^2d\omega_{D-1}^2
\end{equation}
It follows that along Minkowski causal curves, we have
\begin{equation}
    \begin{split}
    \dot{t}^2-\dot{r}^2&\geq0\\
       \implies \dot{v}=0\text{ or }\frac{du}{dv}&=\frac{\dot{t}-\dot{r}}{\dot{t}+\dot{r}}\geq0
    \end{split}
\end{equation}
where $\dot{}$ denotes differentiation with respect to an affine parameter $s$. Using this, Figure \ref{fig:causaldiamond} illustrates how Lemma \ref{lemma:confinetolargeR} is true in Minkowski spacetime.

\begin{figure}
    \centering
    \includegraphics[scale=0.3]{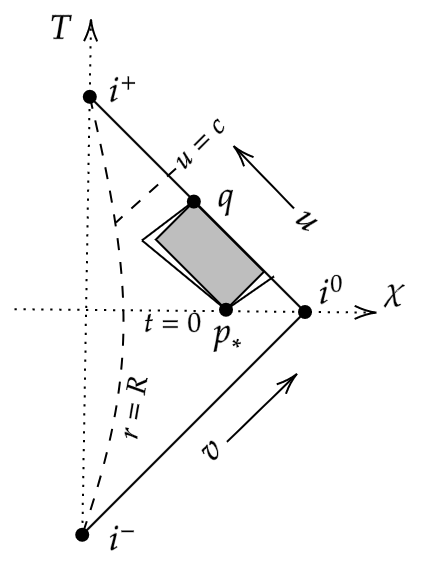}
    \caption{The shaded region shows $J^+(p_*)\cap J^-(q)$ calculated with respect to the Minkowski metric. This region is bound by curves with $\left|\frac{dT}{d\chi}\right|=1$ For any $c$ and $\epsilon>0$, $g$-causal curves in $\{u<c\}\cap\{r>R_0\}\cap\{t\geq0\}$, with $R_0$ sufficiently large, have $\left|\frac{dT}{d\chi}\right|\geq1-\epsilon$.}
    \label{fig:causaldiamondPMT}
\end{figure}

The need to prove this lemma is the downside of compactifying using $u=t-r$ and $v=t-r$. It is possible that $g$--causal curves from $p_*$ have $\frac{du}{dv}<0$ and hence escape the shaded region shown in Figure 4. The proof below relies on results obtained in the proof of Lemma \ref{lemma:compactification}. These are used to show that $g$-null geodesics and $\eta$-null geodesics emanating from the same point remain close in a neighbourhood of $i^0$ (see Figure \ref{fig:causaldiamondPMT}). 

\textbf{Proof:} Let $\gamma$ be a $g$--null geodesic from $p_*$ to $q$ and let $s$ be an affine parameter along $\gamma$ (increasing to the future) such that $\frac{dx^i}{ds}\frac{dx^i}{ds}|_{s=0}=1$ and $\frac{dr^2}{ds}|_{s=0}\geq0$. Let $s_0$ denote the value of $s$ at $p_*$.

From Figure \ref{fig:causaldiamond} we see that, for fixed $R$, in order for a curve from $p_*$ to $q$ (with $r$ sufficiently large at $p_*$) to enter the region $r<R$, it must have $\frac{du}{dv}<0$ at some point.

We first consider the case where $\dot{u}<0<\dot{v}$ along some portion of $\gamma$. It is possible, if $\gamma$ reaches small values of $r$, that we may have $\dot{t}\leq0$ at some point. However since the metric is asymptotically flat, there exists $\delta>0$ such that $\dot{t}>\delta>0$ for sufficiently large $r$. We therefore have $\dot{v}=\dot{t}+\dot{r}>2\dot{t}>2\delta$.

It follows that for $r>R_0$ (with $R_0$ sufficiently large) we have
\begin{equation}\label{eqn:dudv}
\begin{split}
    0<-(\dot{t}-\dot{r})(\dot{t}+\dot{r})&\leq\eta_{\mu\nu}\dot{x}^\mu\dot{x}^\nu\\
    &\leq(\eta_{\mu\nu}-g_{\mu\nu})\dot{x}^\mu\dot{x}^\nu\\
    &\leq Br^{-(D-2)}\\
    \implies-B'r^{-(D-2)}\leq\frac{du}{dv}=\frac{\dot{t}-\dot{r}}{\dot{t}+\dot{r}}&<0
\end{split}
\end{equation}
for some constants $B,B'>0$.

Choose $c$ such that the $u$ co-ordinate of $q$ satisfies $u_q<c$. Consider a segment of $\gamma$ in the region $\{u<c\}\cap\{r>R_0\}\cap\{t\geq0\}$. We have 
\begin{equation}
\begin{split}
    0\leq t&<r+c\\
    \implies 0\leq v&< 2r+c
\end{split}
\end{equation}

From this we have
\begin{equation}
\begin{split}
     \left|\frac{1+v^2}{1+u^2}\frac{du}{dv}\right|&\leq \left(1+(2r+c)^2\right)\frac{B'}{r^{D-2}}\\
     &\longrightarrow0 \text{ as }r\longrightarrow\infty
\end{split}
\end{equation}
Let $\epsilon>0$. From the above we see that for $R_0$ sufficiently large we have
\begin{equation}
    1-\epsilon\leq\frac{dT}{d\chi}=\frac{1+\frac{1+v^2}{1+u^2}\frac{du}{dv}}{1-\frac{1+v^2}{1+u^2}\frac{du}{dv}}\leq1
\end{equation}
for $r>R_0$, where $T$ and $\chi$ are defined in (\ref{eqn:Tchi}).

Now consider the case where $\dot{v}<0<\dot{u}$ along a portion of $\gamma$. We have $\dot{u}=\dot{t}-\dot{r}>2\dot{t}>2\delta$ for $r$ sufficiently large. Analogously to (\ref{eqn:dudv}), it follows that for $r>R_0$ (with $R_0$ sufficiently large) we have
\begin{equation}
\begin{split}
0&>\frac{du}{dv}=\frac{\dot{t}-\dot{r}}{\dot{t}+\dot{r}}>-B''r^{-(d-2)}
\end{split}
\end{equation}
for some constant $B''>0$.

We also have
\begin{equation}
\begin{split}
     \left|\frac{1+u^2}{1+v^2}\frac{dv}{du}\right|&\leq \frac{B''}{r^{d-2}}\\
     &\longrightarrow0 \text{ as }r\longrightarrow\infty
\end{split}
\end{equation}

Hence for any $\epsilon>0$, if $R_0$ is sufficiently large we have
\begin{equation}
    -1\leq\frac{dT}{d\chi}=\frac{\frac{1+u^2}{1+v^2}\frac{dv}{du}+1}{\frac{1+u^2}{1+v^2}\frac{dv}{du}-1}\leq-1+\epsilon
\end{equation}
in the region $\{u<c\}\cap\{r>R_0\}\cap\{t\geq0\}$.

Putting these results together, we find that for any $c$ and any $\epsilon>0$, there exists some $R_0>0$ such that causal curves in the region $\{u<c\}\cap\{r>R_0\}\cap\{t\geq0\}$ have 
\begin{equation}
  \left|\frac{dT}{d\chi}\right|\geq1-\epsilon.
\end{equation}
From this we conclude that $g$-causal curves in this region are confined to a wedge in the Penrose diagram bound by straight lines with gradients which, by choosing $R_0$ sufficiently large, can be chosen arbitrarily close to $\pm1$. It follows that for any $R>0$, there exists some $R'$ such that if $p_*$ has $r$ co-ordinate $r_{p_*}>R'$ then any $g$-causal curve from $p_*$ to $q$ does not enter the region $r\leq R$. This is illustrated in Figure \ref{fig:causaldiamondPMT}. \qedsymbol

\textbf{Theorem \ref{thm:nullline}} Let $(M,g)$ be a uniformly Schwarzschildean spacetime in $D+1$ dimensions ($D\geq4$) with ADM mass $m_{ADM}<0$ and suppose $\mathcal{D}\cup\mathcal{I}$ is a globally hyperbolic subset of $\tilde{M}$. Then $(M,g)$ contains a null line.

\textbf{Proof:} Using Lemma \ref{lemma:negativetimeofflight}, we construct a causal curve from $\Lambda^-$ to $\Lambda^+$ which has negative time of flight. Denote this curve by $\gamma_0$ and label its past and future endpoints $p_0$ and $q_0$ respectively. 

Following \cite{PSW}, we introduce a partial ordering $\leq$ on $\left(\left(J^+(p_0)\cap \Lambda^-\right)\cup i^0\right)\times\left(\left(J^-(q_0)\cap \Lambda^+\right)\cup i^0\right)$. We say that $(p',q')\leq (p'',q'')$ if $p'\in J^+(p'')$ and $q'\in J^-(q'')$. Define the set $F$ to contain all pairs of points in $\left(J^+(p_0)\cap \Lambda^-\right)\times\left(J^-(q_0)\cap \Lambda^+\right)$ which can be connected by a causal curve through $\mathcal{D}$ and choose $(p,q)$ to be any element of the closure of this set which is minimal with respect to the partial ordering $\leq$ (so a priori one or both of these points could lie at $i^0$).

We can then define sequences of points $\{p_i\}_{i=0}^\infty$ along $\Lambda^-$ and $\{q_i\}_{i=0}^\infty$ along $\Lambda^+$ such that
\begin{itemize}
    \item $(p_i,q_i)\leq(p_j,q_j)$ for $i>j$
    \item $p_i$ and $q_i$ are connected by a causal curve $\gamma_i$
    \item $p_i\longrightarrow p$ and $q_i\longrightarrow q$ as $i\longrightarrow\infty$
\end{itemize}
The first condition here ensures that the time of flight along $\gamma_i$ is less than or equal to the time of flight along $\gamma_j$ for $j<i$.


We first check that the sequence of curves $\{\gamma_i\}_{i=0}^\infty$ does not escape to infinity. By this we mean that there exists some $R>0$ such that, possibly restricting to a subsequence, every curve $\gamma_i$ enters the region $\{r<R\}$. 

The curve $\gamma_i$ can be replaced by a (possibly faster) curve $\gamma_i'$ from $\Lambda^-$ to $\Lambda^+$ which is the union of two null geodesics joined at $p_{*,i}$, the point at which $\gamma_i$ intersects the surface $t=0$. 

Let $R_i$ denote the value of the $r$ co-ordinate at $p_{*,i}$. Suppose that $R_i\longrightarrow\infty$ as $i\longrightarrow\infty$. From Lemma \ref{lemma:confinetolargeR}, for any $R>0$ (and restricting to a subsequence if necessary) the sequence of curves $\{\gamma_i\}_{i=0}^\infty$ is eventually contained in the region $r>R$. 

Then, by Lemma \ref{lemma:timeofflight}, we have that the time of flight along $\gamma_i'$ satisfies
\begin{equation}
    |u_{\gamma_i',\infty}-v_{\gamma_i',\infty}|\leq \frac{A}{R^{D-3}}
\end{equation}
for some constant $A$, where we note that a uniformly Schwarzschildean spacetime admits quasi-Cartesian co-ordinates with $\alpha=D-2$.

From this we see that, for $R$ sufficiently large, the time of flight along the $\gamma_i'$, and hence also the time of flight along $\gamma_i$, will eventually become strictly greater than the time of flight along $\gamma_0$ (which was chosen to be strictly negative). But this contradicts the definition of the sequence $(\gamma_i)_{i=0}^\infty$ as consisting of faster and faster causal curves. We therefore conclude that, restricting to a subsequence if necessary, each of the causal curves $\gamma_i$ enters the set $\mathcal{K}:=\left(J^+(p_0)\cap J^-(q_0)\right)\setminus \mathcal{U}_R$, where $\mathcal{U}_R:=\{x\in J^+(p_0)\cap J^-(q_0):r(x)>R\}$ and $J^+(p_0)\cap J^-(q_0)$ includes $i^0$ as well as certain points on $\Lambda^+\cup\Lambda^-$.

But $\mathcal{U}_R$ is an open set in $J^+(p_0)\cap J^-(q_0)$ and, since $\mathcal{D}$ is globally hyperbolic, $J^+(p_0)\cap J^-(q_0)$ is compact. It follows that $\mathcal{K}$ is a compact set. Hence, defining $\gamma$ to be the limit of the sequence $\{\gamma_i\}_{i=0}^\infty$ in the Vietoris topology (Appendix A, \cite{SorkinWoolgar}), we see that $\gamma$ must also enter $\mathcal{K}$. In particular this means that $\gamma$ enters the interior of the spacetime. Furthermore, since each of the $\gamma_i$ are causal, so too is the limit curve $\gamma$.

So $\gamma$ is a fastest causal curve from $\Lambda^-$ to $\Lambda^+$ which enters the interior of the spacetime. Such a curve necessarily lies on the boundary of the future null cone from $p$ and hence must be a null geodesic (\cite{Wald} Corollary after Theorem 8.1.2) without conjugate points (\cite{HawkingEllis} Proposition 4.5.12). We conclude that $\gamma$ is a null line.
\section{A Focusing Theorem in Higher Dimensions}\label{A Focusing Theorem in Higher Dimensions}
Theorem \ref{thm:nullline} shows that certain higher dimensional, negative mass spacetimes contain a null line.  This means that such spacetimes would be excluded if we also impose conditions which forbid the existence of such a curve.

We will require conditions which ensure that endless null geodesics encounter sufficient regions of positive focusing to guarantee that conjugate points will occur. This focusing of geodesics is consistent with the sort of behaviour we would expect to be caused by regions of positive mass (assuming the Einstein equations hold). In this sense, Corollary \ref{cor} below agrees with our prior understanding of positivity of mass.

In \cite{PSW}, the conditions imposed are those stated in Borde's Focusing Theorem \cite{Borde} (where we require these to hold for every complete causal geodesic). As mentioned in Section \ref{Introduction}, we now assume that the quantity $R_{ab}T^aT^b$ is finite and continuous, where $T^a$ is tangent to the null geodesic under consideration. To guarantee this, it is sufficient to assume that the metric is $C^2$. 
\begin{thm}(\cite[Focusing Theorem 2]{Borde})\label{thm:Borde}
Let $\gamma$ be a complete, affinely parameterised, causal geodesic with tangent $T^a$ such that $T_{[a}R_{b]cd[e}T_{f]}T^cT^d\neq0$ at some point on $\gamma$. Suppose that for any $\epsilon>0$ and any $t_1<t_2$, there exists $\delta>0$ and intervals $I_1$ and $I_2$ of length $\geq\delta$ with endpoints $<t_1$ and $>t_2$ respectively such that  
\begin{equation}
    \int_{t'}^{t''}R_{ab}T^aT^bdt\geq-\epsilon \quad\quad\forall t'\in I_1,\quad \forall t''\in I_2
\end{equation}
Then $\gamma$ contains a pair of conjugate points.
\end{thm}

Theorem \ref{thm:Borde} was originally stated in $3+1$ dimensions, although the proof does not rely on this and hence the theorem also holds in higher dimensions. For this reason, we may use the conditions of this theorem in our higher dimensional generalisation of the positive mass theorem. As is pointed out in \cite{PSW}, if the Einstein equations are assumed to hold then the conditions of Borde's theorem can equivalently be expressed in terms of the energy-momentum tensor. 

Note that the conditions of Borde's theorem are entirely global. Other conditions often used in such focusing theorems relate to local positivity of energy (assuming the Einstein equations hold). These conditions can be violated in a quantum theory, so our approach has the advantage that it may be possible to extend it to the semi-classical regime. 

It may be the case that we could impose weaker conditions than those used in Borde's theorem and still rule out the existence of null lines. For example, Theorem \ref{thm:nullline} only required the metric to be $C^{1,1}$, so a focusing result for metrics which fail to be $C^2$ would provide a more general result. This is mentioned at the end of Section II.2 in \cite{PSW} and also in \cite{Perspectives}. The important thing is that we impose sufficiently strong conditions to ensure that null lines cannot exist. The condition that $T_{[a}R_{b]cd[e}T_{f]}T^cT^d\neq0$ at some point on every geodesic is called the \textit{generic condition} and is satisfied in all but a very special class of spacetimes \cite{PenHaw}. As discussed in \cite{PSW}, it does not appear to be a particularly restrictive assumption. This is because if it is not satisfied by our spacetime, then we would expect it to be satisfied by some other spacetime which is ``nearby'' in some sense and whose 4-momentum differs by an arbitrarily small amount. As a result, imposing this assumption does not appear to weaken the positivity of mass result obtained here. 

Using Theorems \ref{thm:nullline} and \ref{thm:Borde}, we derive the following corollary, which is our version of the positive mass theorem in higher dimensions.
\begin{cor}\label{cor}
Let $(M,g)$ be a uniformly Schwarzschildean spacetime in $D+1$ dimensions ($D\geq4$) which satisfies the conditions of Borde's theorem for every complete causal geodesic and is such that $\mathcal{D}\cup\mathcal{I}$ is globally hyperbolic as a subset of $\tilde{M}$. Then $m_{ADM}\geq0$.
\end{cor}

\section{Conclusion}\label{Conclusion}
We have proved a version of the positive mass theorem for higher dimensional uniformly Schwarzschildean spacetimes. As mentioned in Section \ref{Definitions and Assumptions}, the assumptions made about our spacetime are weaker than those of \cite{poorman2}. In particular, we only require the metric to be uniformly Schwarzschildean, whereas in \cite{poorman2} the more restrictive requirement that the metric be \textit{strongly uniformly Schwarzschildean} is made. We also drop the assumption of weak asymptotic regularity. However, as in \cite{poorman2} we fail to prove anything regarding the $m=0$ case (a similar problem is encountered in \cite{PSW} in 3+1 dimensions). This is because the method used in Section \ref{Constructing a Fastest Causal Curve} to construct the initial timelike curve $\gamma_0$ with strictly negative time of flight is no longer guaranteed to work. Its success will be determined by the higher order terms in the metric. 

The assumptions made in this paper are also fundamentally different to those used in the proofs by Witten \cite{Witten} and Schoen and Yau \cite{SchoenYau1,SchoenYau2,SchoenYau3} (extended to all dimensions up to seven by Eichmair, Huang, Lee and Schoen \cite{EHLS}). The proof given here relies on properties of the spacetime in a neighbourhood of conformal infinity, whereas these other methods have the advantage of imposing conditions only on some initial spacelike hypersurface. Nonetheless, the ``global'' proof presented in this paper is interesting because it relies more clearly on properties we expect to hold in positive mass spacetimes. In particular, we show that null geodesic focusing is compatible only with spacetimes of non-negative mass. Consequently this proof also acts as evidence that it is consistent to think of the ADM mass defined in higher dimensions as describing the total mass contained in a spacetime. Furthermore, as mentioned above and in \cite{PSW}, by relying on global focusing results this proof avoids imposing non-negativity of energy locally. This leads to the possibility that the methods described here can be generalised to the semi-classical setting, where such local conditions can be violated by quantum matter. 
\subsection*{Acknowledgements} I am grateful to Maciej Dunajski, Piotr Chru\'{s}ciel, Greg Galloway, Roger Penrose, Harvey Reall, Paul Tod, Eric Woolgar, Gary Gibbons and Claude Warnick for useful discussions, as well as to the anonymous referee for their extremely helpful comments. I am also grateful to St. John's College, Cambridge for their support through a College Scholarship from the Todd/Goddard Fund. Some of this work was carried out while I was visiting Institut Mittag-Leffler in Djursholm, Sweden, where I was supported by the Swedish Research Council under grant no. 2016-06596.
\begin{appendices}
\section{Compactness of Causal Curves Following \cite{SorkinWoolgar}}
In this appendix we summarise how the convergence of the sequence of causal curves $(\gamma_i)_{i=0}^\infty$ is defined. We define $\gamma$ to be the limit of this sequence in the Vietoris topology \cite{SorkinWoolgar}. To understand this convergence, we follow \cite{Convergence} and make the following definitions:
\begin{definition}
The sequence of causal curves $(\gamma_i)_{i=0}^\infty$ converges to $\gamma$ in the \textit{upper topology} iff the $\gamma_i$ are eventually included in every open set which includes $\gamma$.
\end{definition}
\begin{definition}
The sequence of causal curves $(\gamma_i)_{i=0}^\infty$ converges to $\gamma$ in the \textit{lower topology} iff the $\gamma_i$ eventually meet every open set which meets $\gamma$.
\end{definition}
\begin{definition}
The sequence of causal curves $(\gamma_i)_{i=0}^\infty$ converges to $\gamma$ in the \textit{Vietoris topology} iff it converges in both the upper and lower topologies.
\end{definition}
Since we have assumed that $\mathcal{D}\cup\mathcal{I}$ is globally hyperbolic, if we were also to assume that the metric were $C^2$ (as in Theorem \ref{thm:Borde}) then these three definitions would coincide (\cite{SorkinWoolgar} Alternative Definition 26 and comments after proof of Lemma 28). 
\end{appendices}

\end{document}